# Erbium doped yttrium oxide thin films grown by chemical vapour deposition for quantum technologies

Anna Blin[1], Alexander Kolar[2], Andrew Kamen[2], Qian Lin[2], Xiaogang Liu[2], Aziz Benamrouche[3], Romain Bachelet[3], Philippe Goldner[1], Tian Zhong[2], Diana Serrano[1*] and Alexandre Tallaire[1#]

[1] IRCP, CNRS, PSL Research University, 11 rue Pierre et Marie Curie, 75005, Paris, France

[2] Pritzker School of Molecular Engineering, University of Chicago, Chicago, Illinois 60637, USA

[3] INL, Université de Lyon, Ecole Centrale de Lyon, CNRS UMR 5270, 69134 Ecully, France

* Corresponding author: diana.serrano@chimieparistech.psl.eu

# Corresponding author: alexandre.tallaire@chimieparistech.psl.eu

**Abstract**

The obtention of quantum-grade rare-earth doped oxide thin films that can be integrated with optical cavities and microwave resonators is of great interest for the development of scalable quantum devices. Among the different growth methods, Chemical Vapour Deposition (CVD) offers high flexibility and has demonstrated the ability to produce oxide films hosting rare-earth ions with narrow linewidths. However, growing epitaxial films directly on silicon is challenging by CVD due to a native amorphous oxide layer formation at the interface. In this manuscript, we investigate the CVD growth of erbium-doped yttrium oxide (Er:$Y_2O_3$) thin films on different substrates, including silicon, sapphire, quartz or yttria stabilized zirconia (YSZ). Alternatively, growth was also attempted on an epitaxial $Y_2O_3$ template layer on Si (111) prepared by molecular beam epitaxy (MBE) in order to circumvent the issue of the amorphous interlayer. We found that the substrate impacts the film morphology and the crystalline orientations, with different textures observed for the CVD film on the MBE-oxide/Si template (111) and epitaxial growth on YSZ (001). In terms of optical properties, $Er^{3+}$ ions exhibit visible and IR emission features that are comparable for all samples, indicating a high-quality local crystalline environment regardless of the substrate. Our approach opens interesting prospects to integrate such films into scalable devices for optical quantum technologies.

## 1. Introduction

The research for spectrally narrow, bright and stable quantum emitters considers a variety of solid-state platforms, including defect centres in wideband semiconductors such as diamond or silicon carbide (SiC), defects in Si, quantum dots and rare-earth ions in crystals [1–4]. The emitter's transition wavelength is also of utmost importance and a lot of efforts has been devoted to identifying quantum emitters compatible with optical fibre networks. In this context, trivalent erbium ions ($Er^{3+}$) are an ideal candidate as they present an optical transition in the telecom band around 1.5 µm and narrow spectral linewidth down to 73 Hz when inserted into high-quality low-magnetic-density crystals [5]. As a drawback, $Er^{3+}$ presents a low photon emission rate at this wavelength, of the order of 100 photons per second, and therefore requires taking advantage of photonic enhancement for observing single ions. In recent years, single $Er^{3+}$ ions have been detected using different architectures, including oxide nanoparticles and oxide thin films coupled to fibre cavities [6,7], as well as Si photonic resonators deposited on high-quality bulk oxide crystals [2,8]. The development of nanoscale rare-earth doped materials with good coherence properties is therefore crucial to successfully achieve their integration into such quantum devices. Epitaxial oxide thin films doped with $Er^{3+}$ and deposited on scalable and easily processable wafers such as Si stand out as they would offer simultaneous integration with photonic and MW cavities [7].

However, deposition of functional crystalline oxide films on semiconductors is quite challenging essentially because of their dissimilarity with the substrate in terms of lattice parameter or thermal mismatch. In addition, unwanted oxidation of the substrate can lead to an amorphous interfacial $SiO_2$ layer on Si in the initial growth stage, which prevents further epitaxial deposition of the film [9]. Several demonstrations have nevertheless been made, with yttria stabilized zirconia (YSZ) and strontium titanate (STO) on Si being the most popular routes so far. The approach is based on a thorough control of the oxidation of the semiconductor surface and specific interface engineering. In this sense, molecular beam epitaxy (MBE) [10] as well as pulsed laser deposition (PLD) [11] are the most adapted techniques to achieve oxide epitaxy on Si. They offer a good control of the interface thanks to the use of very low pressures that can prevent formation of unwanted oxides and advanced in-situ analysis techniques to finely tune the composition while growing. However, chemical vapour deposition potentially offers better prospects for upscaling, higher flexibility and increased growth rates at a reduced cost. It would therefore be desirable to develop a CVD process that allows growing high quality or even epitaxial oxides containing rare-earth ions that are suitable for the foreseen use in quantum technologies.

In this study, we synthesize Er-doped $Y_2O_3$ thin films by direct liquid injection CVD on different substrates and assess their optical properties in view of their use for quantum technologies. We aim at maximizing their crystallinity and explored two main routes: (i) films were either grown on bulk commercial oxide single crystals as they offer ideal chemical compatibility and better prospects for epitaxy, alternatively (ii) films were grown on Si substrates onto which a thin epitaxial $Y_2O_3$ buffer layer was previously deposited by MBE. This hybrid integration provides a way to promote epitaxy of the "active" overlayer containing the rare-earth ions on this scalable substrate. As a comparison, $Er^{3+}$:$Y_2O_3$ films directly grown by CVD on (111) and (100) Si substrates were also prepared. The structural and optical properties of the films were eventually assessed and benchmarked to bulk ceramics to get a better insight into their relevance. We found that all substrates led to well-crystallized thin films with comparable emission properties for $Er^{3+}$ ions; however, from the morphological point of view, two substrates stand out: YSZ, onto which epitaxial growth is achieved, and the MBE template oxide on silicon, featuring high texture and the largest crystalline domains among all samples. Low temperature high resolution spectroscopic investigations of the telecom transition of $Er^{3+}$ in these two substrates showed narrow inhomogeneous linewidths varying between 9 and 14 GHz for ions in the $C_2$ site.

## 2. Experimental methods

The yttria films were deposited by direct liquid injection CVD using a home-made reactor described in previous papers [12,13]. The yttrium and erbium precursors were Y(tmhd)$_3$ and Er(tmhd)$_3$ (where "tmhd" accounts for tetra-methyl-heptane-dionate). They were dissolved in mesitylene and flown into a 2-injection head vaporization box heated to 210 °C (*Vapbox 1500* from *Kemstream*), where they were flash-evaporated [14]. The deposition was performed at a temperature of about 850 °C, with 30 % $O_2$ in $N_2$ at a total pressure of about 13 mbar. Different 10x10 mm² commercial substrates were used for the film deposition, as indicated in the table.

| Reference | Substrate | Film | Film thickness (nm)* | Lattice mismatch at 850°C (%) | Domain matching | Thermal mismatch (%) |
|---|---|---|---|---|---|---|
| AB-2204 | Si (100) | Er:$Y_2O_3$ | 272 nm | -1.8 | 2 | 200 |
| AB-2302 | Si (111) | Er:$Y_2O_3$ | 229 nm | -1.8 | 2 | 200 |
| AB-2321 | Si (111) / $Y_2O_3$-MBE | Er:$Y_2O_3$ | 60 nm + ~250 nm | -1.8 | 2 | 200 |
| AB-2306 | Z-cut Quartz (0001) | Er:$Y_2O_3$ | 252 nm | 1.2 | 3 | 5 |
| AB-2304 | Sapphire (0001) | Er:$Y_2O_3$ | 229 nm | 4.8 | 3 | 40 |
| AB-2303 | YSZ (001) | Er:$Y_2O_3$ | 229 nm | 2.8 | 2 | -15 |

*Table 1. Summary of the samples studied in this work. Films were doped with a nominal concentration of $Er^{3+}$ of 670 ppm. The domain matching represents the number of unit cells of the substrate with respect to that of the $Y_2O_3$ film. Thermal mismatch accounts for the difference in thermal expansion coefficients between the film and the substrate. Note that sample thickness was estimated from ellipsometric measurement performed on a film grown on a silicon substrate simultaneously.*

The $Y_2O_3$ template was grown epitaxially on a silicon (111) wafer using MBE at temperatures between 600 and 920 °C. Deposition was monitored in-situ during growth using reflection high electron diffraction (RHEED) as described in [15]. The film thickness was around 60 nm.

Images of the surface of the films were taken with a scanning electron microscope with a field emission gun (FEG-SEM) *ZEISS Merlin*. To identify the favoured growing directions of the films, an X-ray diffractometer *PANalytical XPert Pro* with a monochromatic Cu-K$\alpha_1$ radiation was used. A high-brilliance X-ray diffractometer (*SmartLab*) from *Rigaku*, equipped with a 9 kW rotating anode and a two-bounce Ge (220) monochromator was also used to further investigate the crystalline orientation and the structure of the Er-doped $Y_2O_3$ films on YSZ. Atomic force microscopy (AFM, with a *NX10* equipment from *Park Systems*) was used in tapping mode with a conical tip (radius < 5 nm, full angle 30°) for topographic analyses.

The optical properties of $Er^{3+}:Y_2O_3$ films on their substrates were measured using a tunable optical parametric oscillator (OPO) pumped by a $Nd^{3+}$ YAG Q-switched laser (*Ekspla NT342B-SH* with 6 ns pulse length and 10 Hz repetition rate). Spectra were recorded by a spectrometer (*Acton SP2300*) equipped with several holographic gratings (300 grooves/mm and 1200 grooves/mm for visible range and 600 grooves/mm for IR range) and an ICCD camera (*Princeton Instruments*), while a photomultiplier tube (*Hamamatsu*) and an InGaAs photodiode (*Femto OE-200-IN1*) were used as detectors for lifetime measurements in the visible and IR ranges respectively. The samples were mounted on the cold finger of a closed cycle helium cryostat and cooled down to ≈ 10 K.

To probe the IR transitions with a higher resolution, the samples were mounted on a 2.9 K close-cycle cryostat (*FourNineDesign*). A continuously tunable diode laser (*Toptica CTL1500*) was pulsed using 3 acousto-optic modulators (AOM) in series with a high extinction ratio of 160 dB while its wavelength was monitored. Photoluminescence from $Er^{3+}$ was detected by a superconducting nanowire single photon detector (SNSPD) with ≈70 % detection efficiency and a 2.1 count per second (cps) dark count rate.

### 3. Structural analysis of the films

The surface morphology of the $Er^{3+}:Y_2O_3$ thin films grown on various substrates and observed by SEM is shown in Fig. 1 for comparison. Crystalline faceted grains always formed, regardless of the substrate used. Deposition on Si (100), Si (111), quartz and sapphire led to a polycrystalline film with visibly randomly oriented grains having a size in the 50 to 100 nm range. On Si with a MBE-$Y_2O_3$ template layer (Fig. 1c), grains were slightly larger (100-200 nm) but also with no apparent specific shape. On the contrary, when a YSZ substrate was used, a rather different morphology was obtained with well-defined square columns of 30-60 nm in size (Fig. 1e). This is a first hint suggesting that oxide growth proceeds epitaxially on this type of substrate. AFM measurement was then carried out on a $Y_2O_3$ film grown on YSZ to better evaluate the surface morphology over a 0.5x0.5 µm² area (Fig 2a). The square and oriented shape of the grains is confirmed. They exhibit a relatively flat top surface, as shown by the line profiles of Fig. 2b, suggesting that the morphology is essentially composed of (001) facets. The roughness is estimated to Ra = 4.8 nm and RMS = 6.0 nm, which is a relatively high value for such thin films. This rough morphology might essentially come from the 3D growth mode of the film. Further refinement of the growth conditions could help to promote a layer by layer growth mode and potentially a smoother surface [16].

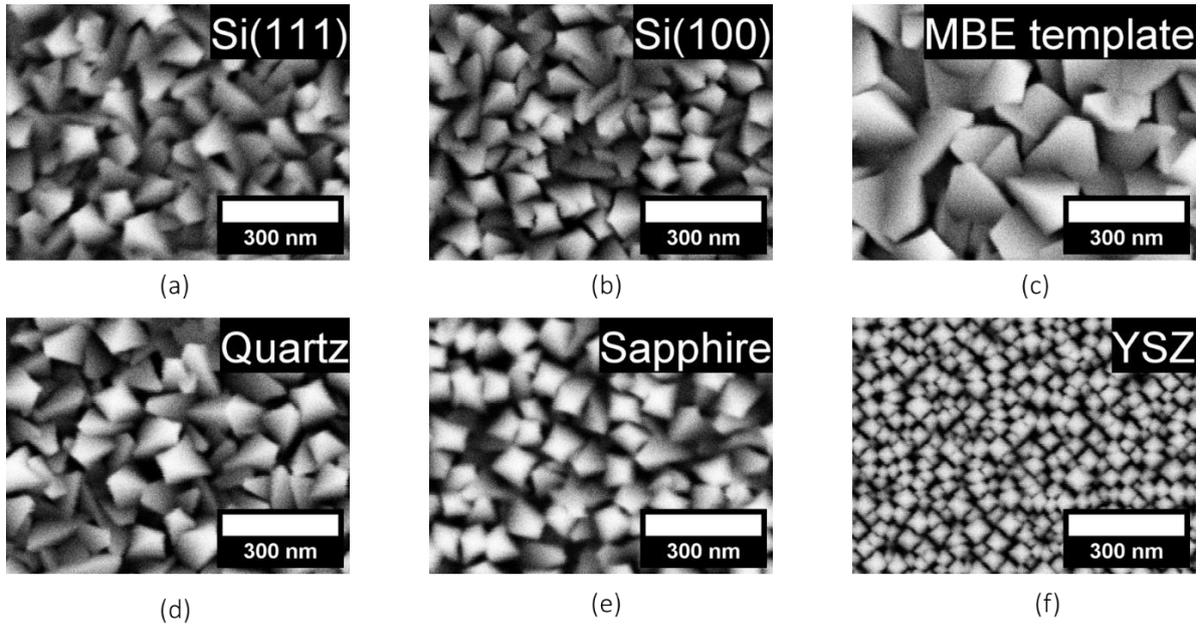

Fig.1. SEM images obtained for Er:Y$_2$O$_3$ ≈ 200 nm-thick films grown by CVD on various substrates. (a) Si (111), (b) Si (100), (c) MBE-grown Y$_2$O$_3$ template on Si (111), (d) (0001) quartz, (e) (0001) c-plane sapphire, (f) (100) yttria stabilized zirconia.

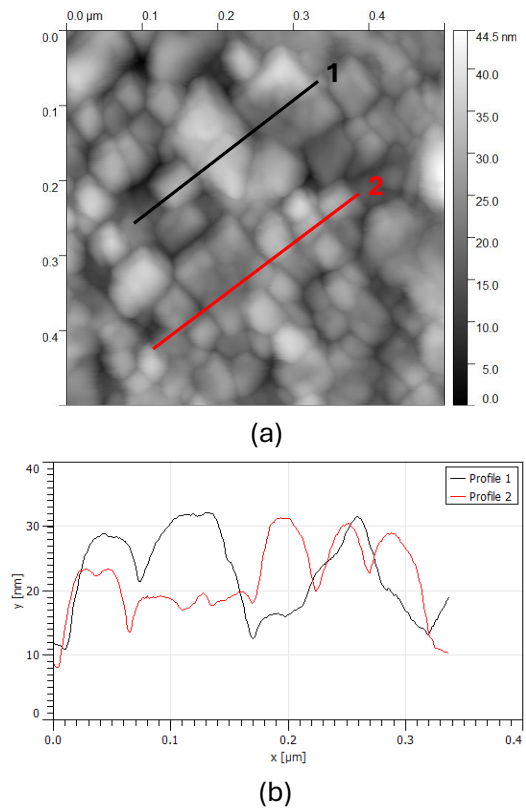

Fig. 2. (a) AFM image of a Y$_2$O$_3$ film grown on YSZ and (b) measured line profiles as indicated on the image.

To get further insight into the texture and orientation of the films, θ-2θ XRD scans were performed. The results are compared to a reference ceramic and presented in Fig. 3a. In the considered angular range, mostly the (222) and (400) diffraction peaks associated with cubic $Y_2O_3$ (space group I-a3) are observed. These peaks are slightly shifted with respect to the ceramic reference, most likely because of thermal mismatch between the film and the substrate. To facilitate observation, a magnification factor indicated above each curve was applied to the (400) peak to increase its relative intensity with respect to that of (222). Furthermore, the following empirical formulae relating the intensity of the (222) and (400) peaks to their expected intensity in a ceramic with randomly oriented grains were used to illustrate the texture (T) formation of the films.

$$T_{(222)}(\%) = 100 \times \left(1 - \frac{\left(I_{(400)}/I_{(222)}\right)_{FILM}}{\left(I_{(400)}/I_{(222)}\right)_{CERAMIC}}\right) \qquad \text{eq. 1}$$

$$\text{or } T_{(400)}(\%) = 100 \times \left(1 - \frac{\left(I_{(222)}/I_{(400)}\right)_{FILM}}{\left(I_{(222)}/I_{(400)}\right)_{CERAMIC}}\right) \qquad \text{eq. 2}$$

The results are presented in Fig. 3b for the (222) and (400) texture in orange and blue. Either formula was used depending on the dominating [111] or [100] texture of the film.

It appears that the films grown on Si (100) and Si (111) are polycrystalline with a rather pronounced [111] texture especially for (111)-oriented Si substrates. Silicon has a strong affinity to oxygen and quickly forms an amorphous oxide layer at its surface. Consequently, the film tends to grow along denser (111) crystal planes without any specific epitaxial relationship with the substrate. On the contrary, when an MBE-grown $Y_2O_3$ template layer on Si (111) is used instead of the uncontrolled amorphous $SiO_2$ layer, a stronger [111] texture (99.4 %) with almost undetectable (400) peak is obtained. Despite this improvement, we were unable to obtain a fully epitaxial film under the conditions that we used, judging by the grain structure observed by SEM (Fig. 1c). Further improvement of the quality of the MBE template layer as well as optimization of the CVD growth conditions would be required to improve epitaxial regrowth.

Regarding oxide substrates, both quartz and sapphire led to the formation of a polycrystalline film with varying degrees of texture. On quartz substrates, the peaks were broader and slightly shifted. On the contrary, when YSZ was used as a substrate, full [100] texture was observed. This indicates that the $Y_2O_3$ film on YSZ grows with out-of-plane orientation identical to that of the substrate. To further resolve the crystalline orientation of the film with respect to the substrate, additional high resolution XRD measurements were acquired. The θ-2θ HR-XRD scan confirms the preferential [100] out-of-plane texture (Fig. 4a), while the rocking curve around this orientation indicates a linewidth of 0.7° (Fig. 4b). This is a reasonable mosaicity for a thin yttrium oxide film [17]. To check the in-plane orientation of the film, XRD analysis was also carried out along off-axis reflections (311) and (611) for YSZ and $Y_2O_3$, respectively (Fig. 4c and 4d). The presence of 2 diffraction peaks in the considered in-plane azimuthal angle range (Δϕ) indicates that the film also has a preferential in-plane orientation. The angular coincidence of these peaks, from these (h11) orientations, supports the fact that growth proceeded epitaxially with a cube-on-cube relationship, with 2 lattice units of the substrate matching one lattice unit of the film.

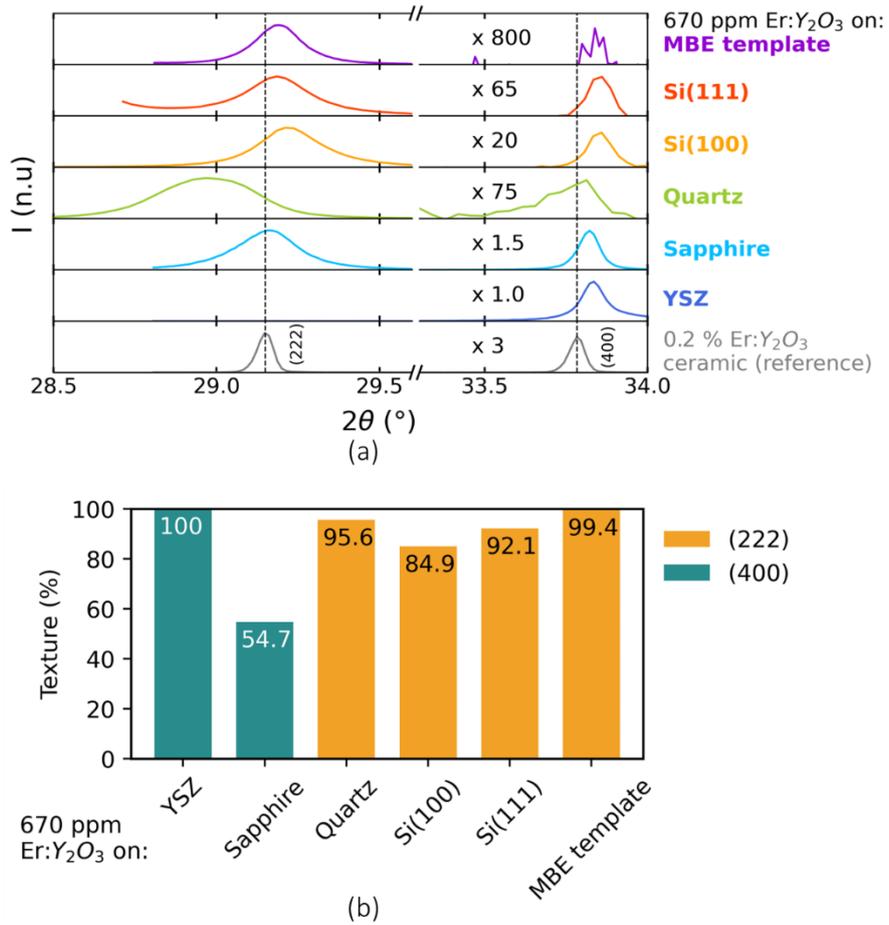

*Fig. 3. (a) θ-2θ XRD scans of the $Y_2O_3$ films grown on different substrates. The position of the (400) and (222) peaks is indicated by dashed lines. The intensity of the (400) diffraction peak has been multiplied by a factor as indicated on the graph for better clarity. (b) Bar diagram showing the calculated texture of the films.*

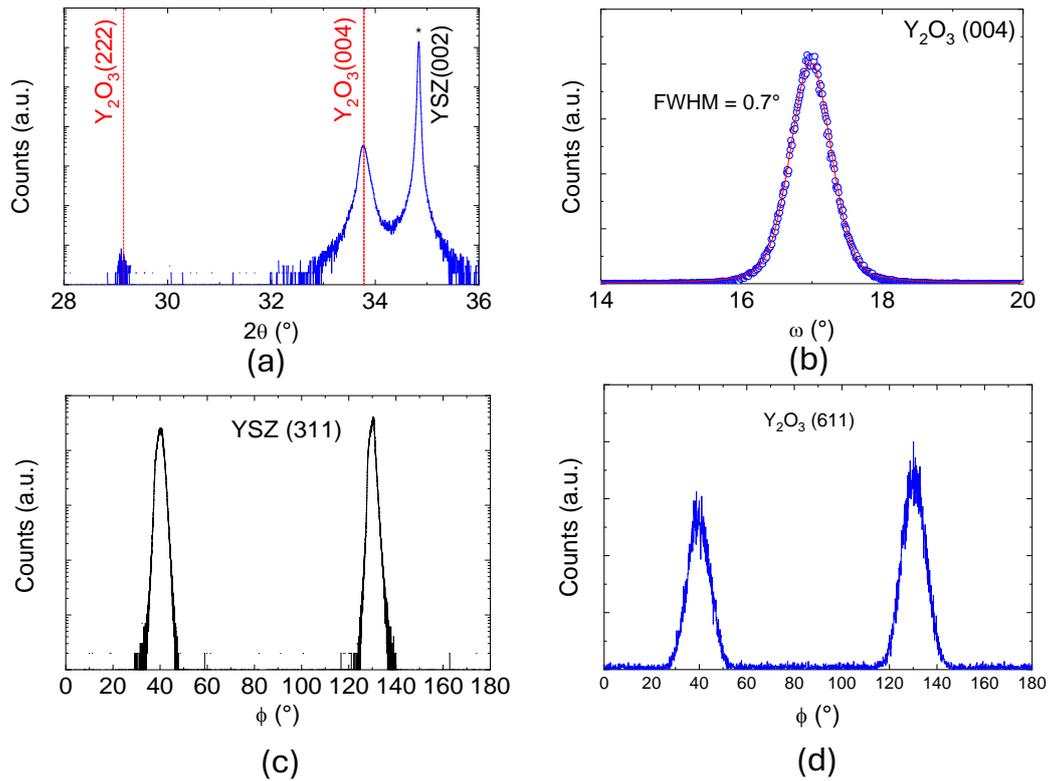

*Fig. 4. (a) High resolution $\theta$-$2\theta$ XRD scans of a $Y_2O_3$ film grown on YSZ. (b) Corresponding omega scan around the (400) peak indicating a 0.7° linewidth. (c) and (d) Phi scans along the (311) and (611) directions of the YSZ substrate and the $Y_2O_3$ film respectively.*

### 4. Spectroscopic analysis of the films

The optical properties of $Er^{3+}$ ions embedded in the $Y_2O_3$ films grown on different substrates were thoroughly evaluated at cryogenic temperatures (10 K). The $^4I_{15/2}$ → $^2H_{11/2}$ transition was resonantly excited at 519.7 nm while PL emission in the visible range was recorded (Fig. 5, ions in the $C_2$ site). Following non-radiative decay from the $^2H_{11/2}$ to the $^4S_{3/2}$ level, several emission lines can be observed between 540 and 640 nm, identified as radiative transitions from the lowest Stark sublevel in the $^4S_{3/2}$ multiplet to the different Stark sublevels in the ground $^4I_{15/2}$ multiplet [18]. All samples show similar PL spectra to that of an $Er^{3+}$-doped $Y_2O_3$ ceramic (grey curve of Fig. 5b). We note that for the quartz substrate, similarly to what had been observed in XRD, the peaks were slightly broadened potentially indicating additional strain or disorder. We conclude that $Er^{3+}$ ions can be found in a high crystalline quality environment in $Y_2O_3$, independently of the substrate used.

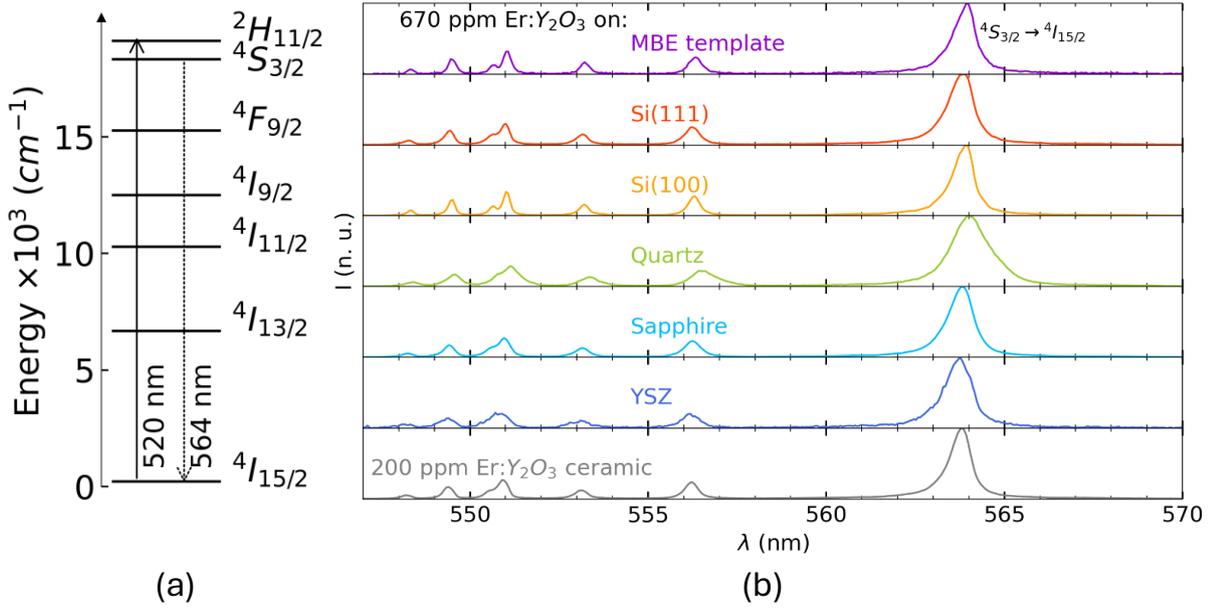

*Fig. 5. Optical properties of the films grown on different substrates. (a) Schematics of the different energy levels of $Er^{3+}$ ions showing the excitation and emission channels (b) Photoluminescence spectra acquired in the visible range at 10 K.*

To get more insight into the close environment of $Er^{3+}$ ions in the films and potential non-radiative decay channels, we focused on the main visible transition at 564 nm corresponding to $^4S_{3/2}$ (0) → $^4I_{15/2}$ (7) as well as the infrared transition at 1536 nm ($^4I_{13/2}$ (0) → $^4I_{15/2}$ (0), (see schematics of Fig. 5a). The IR transition is indeed technologically relevant for telecommunications. Decay curves for films on all substrates are plotted in Fig. 6a and 6b. They were fitted with a bi exponential for the visible $^4S_{3/2}$ (0) → $^4I_{15/2}$ (7) transition and a single exponential for the infrared $^4I_{13/2}$ (0) → $^4I_{15/2}$ (0) transition, to extract characteristic decay lifetimes. Values around 160 µs and 8 ms for the visible and infrared transitions were measured respectively, for ions in the $C_2$ site. Again, these values are reasonably comparable to those measured for a reference ceramic sample and also, in MBE-grown $Er^{3+}$:$Y_2O_3$ thin films [7]. This indicates that the characteristic lifetimes of $Er^{3+}$ levels are preserved in our films regardless of the substrate used. Yet, both visible and IR lifetimes are slightly shorter in the films than in the reference ceramic. This could be due to the presence of non-radiative decay paths, some related to the $Er^{3+}$ doping itself, higher in the thin films than the reference ceramic (670 ppm vs 200 ppm) [18], and to the presence of some uncontrolled luminescence-quenching defects. Surprisingly, for the epitaxial film grown on YSZ, the values are no better or even slightly lower than those of polycrystalline films. This non-intuitive result might be related to the smaller grain size of the film on this substrate leading to PL quenching by defects at grain boundaries. Small differences in the effective $Er^{3+}$ doping level between samples could also partly explain the lifetime differences observed [18].

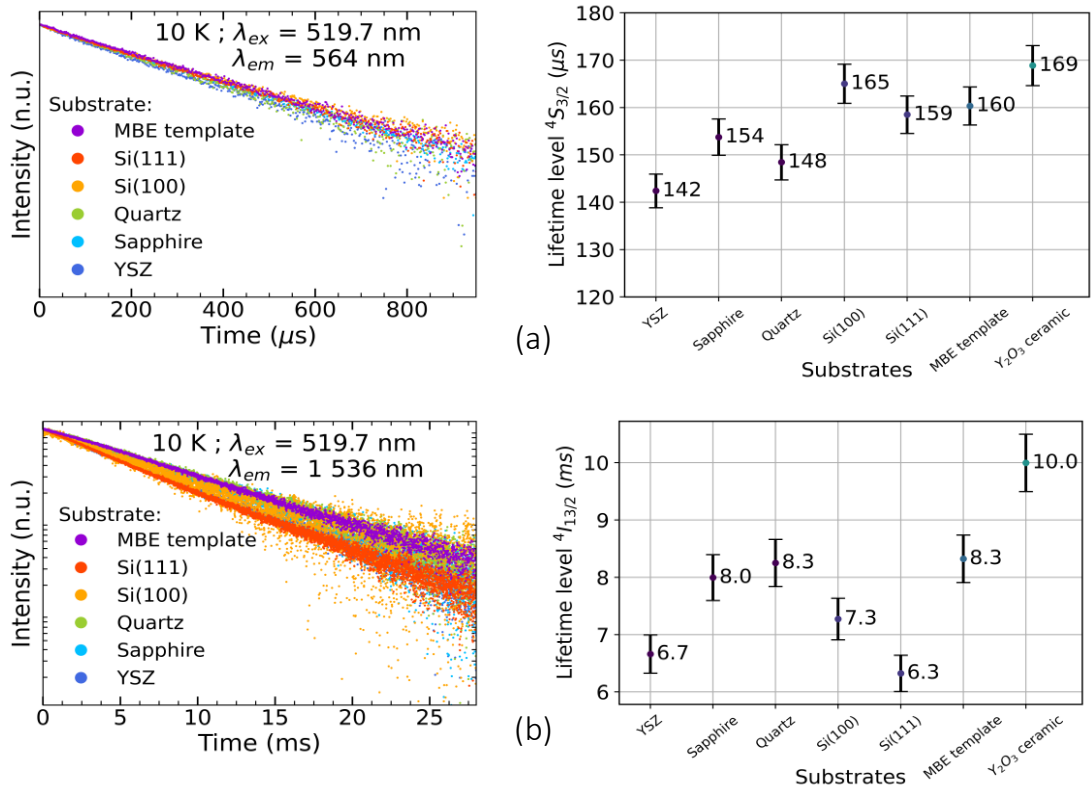

*Fig. 6. PL decay curves of Er$^{3+}$ ions in C$_2$ site acquired (a) in the visible range (564 nm) and (b) in the IR range (1536 nm) at low temperature (10 K) for the Er:Y$_2$O$_3$ thin films grown on different substrates under 519.7 nm excitation. The calculated lifetimes are presented in the right plot for each sample.*

Finally, we focused on the IR transition ($^4I_{13/2} \rightarrow {}^4I_{15/2}$) and performed high-resolution spectroscopy at 2.9 K on the two most interesting systems from the morphological point of view: epitaxial Y$_2$O$_3$ on YSZ and the 99 %-textured Y$_2$O$_3$ on the MBE template. A standard polycrystalline film on Si (100) was also characterized for comparison. At cryogenic temperature, we scanned the laser around 1535 nm and 1545 nm and recorded the PL signal from Er$^{3+}$ ions [19]. These central wavelengths correspond to the 2 crystallographic sites occupied by Er$^{3+}$ ions replacing Y$^{3+}$ in Y$_2$O$_3$ (C$_2$ and C$_{3i}$) [20]. This allows for the assessment of the inhomogeneous broadening ($\Gamma_{inh}$) of the transitions. This linewidth is representative of the static disorder surrounding the ions and is a good measurement of the crystalline quality and purity of the matrix hosting the ions. The obtained values, presented in Fig. 7 and in Table 1, were in the range 9-14 GHz for the C$_2$ site and 16-19 GHz for the C$_{3i}$ site, therefore, rather similar for all three samples. This is of the same order of magnitude as epitaxial MBE grown Er:Y$_2$O$_3$ films [15] but still far from the best values reported in Y$_2$O$_3$ transparent ceramics (0.4 GHz [21,22]). These results also indicate that the local crystalline quality is comparable in polycrystalline thin films and in highly textured or epitaxial ones. This could also mean that the observed spectral inhomogeneity is not directly limited by the film's growth orientation nor morphology, but by other mechanisms such as strain or doping level.

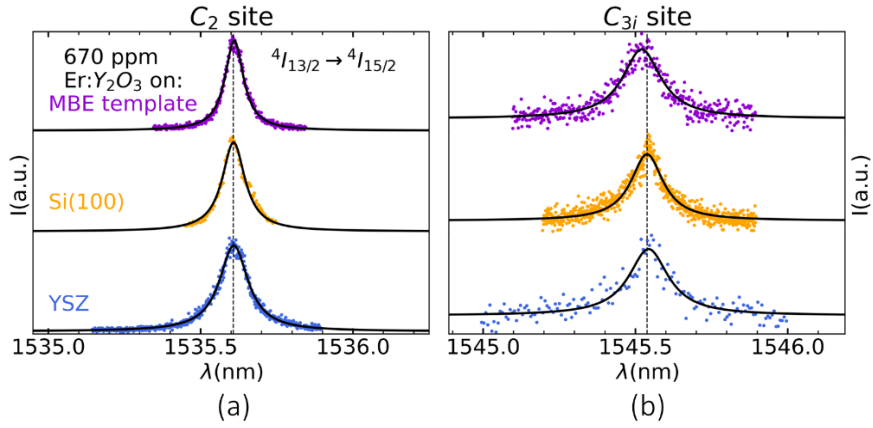

*Fig. 7. Photoluminescence excitation spectra of the films grown on YSZ, Si (100) and Si (111)/$Y_2O_3$-MBE template, showing IR emission for erbium in the $C_2$ (a) and $C_{3i}$ (b) sites as a function of laser excitation wavelength at 2.9 K. The Lorentzian fit associated with the data is plotted as a solid black line.*

| Substrate | $C_2$ (GHz) | $C_{3i}$ (GHz) |
| --- | --- | --- |
| Si (111)/$Y_2O_3$-MBE | 9.4 ± 0.1 | 19.2 ± 0.8 |
| Si (100) | 11.5 ± 0.4 | 16.1 ± 0.4 |
| YSZ (100) | 14.5 ± 0.2 | 19.7 ± 1.6 |

*Table 1. Measured inhomogeneous linewidths derived by Lorentzian fit to the data plotted on Figure 7 of films grown on YSZ, Si (100) and Si (111)/$Y_2O_3$-MBE template.*

To tentatively improve crystalline quality, a post deposition annealing step was applied to the films grown on YSZ and Si (100) substrates. They were heated at 1000 °C in air during 2 h with a heating rate of 200 °C/h and cooled naturally. As for the unannealed samples, high-resolution spectroscopy was performed, and the results are presented in Fig. 8 and Table 2. For both samples, the PL signal is now split into two peaks for the $C_2$ site while it is broadened for ions in the $C_{3i}$ site. This may be due to diffusion of species between the film and the substrate as mentioned in other studies [13] or stress release. On the polycrystalline $Y_2O_3$ film on Si, $C_2$ site inhomogeneous linewidth is about twice larger than on the unannealed sample and the $C_{3i}$ site one, around four times larger. For the film grown on a YSZ substrate, the inhomogeneous linewidth of the $C_{3i}$ site has risen by about 10 GHz while those of the $C_2$ site stays at around 15 GHz.

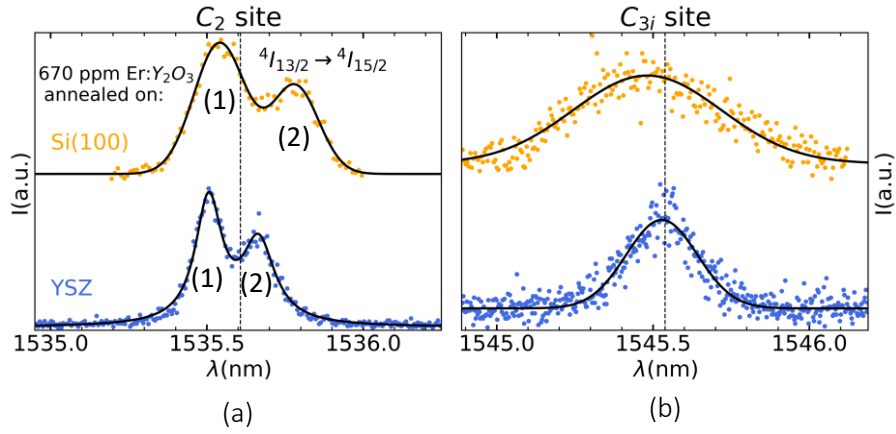

*Fig. 8. Photoluminescence excitation spectra of the films grown on YSZ and Si (100), annealed 2 h at 1000 °C, showing IR emission for Er in the $C_2$ (a) and $C_{3i}$ (b) sites as a function of laser excitation wavelength at 2.9 K. The Lorentzian fit associated with the data is plotted as a solid black line.*

| Substrate | $C_2$ (GHz) peak (1) | $C_2$ (GHz) peak (2) | $C_{3i}$ (GHz) |
| --- | --- | --- | --- |
| Si (100) | 24.7 ± 0.7 | 22.5 ± 0.9 | 70.2 ± 3.2 |
| YSZ (100) | 13.2 ± 0.2 | 16.4 ± 0.3 | 29.3 ± 0.9 |

*Table 2. Measured inhomogeneous linewidths derived by Lorentzian fit to the data plotted on Figure 8 of films grown on YSZ and Si (100) and annealed 2 h at 1000 °C.*

## 5. Conclusions

We have successfully grown high crystallinity Er-doped yttria thin films by DLI-CVD on a variety of substrates, including bare silicon wafers, silicon with a $Y_2O_3$ MBE template oxide layer and bulk single crystal oxides. We observed that the substrate's choice influenced the morphology and texture of the films. We obtained highly (111)-textured films on the $Y_2O_3$-Si MBE template (99 %) and epitaxial (100)-oriented films on YSZ with a 3D morphology composed of square columns. Comparable optical emission from $Er^{3+}$ ions was measured in the visible and IR ranges with typical decay times of about 160 µs and 8 ms for the $^4S_{3/2} \rightarrow {}^4I_{15/2}$ at 564 nm and $^4I_{13/2} \rightarrow {}^4I_{15/2}$ at 1536 nm, respectively. These results were not significantly impacted by the substrate and were comparable to those measured in high quality ceramics with similar doping. Finally, the inhomogeneous broadening of the IR transition from ions in $C_2$ and $C_{3i}$ sites was found to be in the range 9-19 GHz, which is still far from the best values reported in $Y_2O_3$ transparent ceramics. Annealing at high temperature broadened the inhomogeneous linewidth for both sites, the effect being more pronounced for the films grown on Si. This indicates that structural defects need to be reduced in our CVD-grown thin films to improve their optical properties. Future perspective includes the measurement of homogeneous linewidths of $Er^{3+}$ ions, a crucial figure of merit towards their integration into quantum devices such as quantum memories. Efforts will also be devoted to improving their morphology and reducing roughness of the epitaxial $Y_2O_3$ on YSZ which would help designing more efficient waveguides and nanostructures on this platform.

**Ackowledgements:**

The CNRS Prime 80 grant (Hyboxi) and the regional network on quantum technologies of Ile-de-France region (QuanTip) are gratefully acknowledged for funding. This project has also received funding from the European Research Council (ERC) under the European Union's Horizon 2020 research and innovation programme (RareDiamond, grant agreement No 101019234). This project is partially funded by the research collaboration program between the University of Chicago and CNRS, and the France and Chicago Collaborating in The Sciences (FACCTS) program.